\documentclass[aps,prl,showpacs,superscriptaddress,amsmath,amssymb,twocolumn,preprintnumbers]{revtex4}
\usepackage{graphicx}
\usepackage{dcolumn}
\usepackage{bm}
\usepackage{amsmath}
\usepackage{amssymb}
\usepackage[english]{babel}
\usepackage{hyperref}

\newcommand{\beq}{\begin{equation}}
\newcommand{\eeq}{\end{equation}} 
\newcommand{\beqa}{\begin{eqnarray}} 
\newcommand{\eeqa}{\end{eqnarray}}

\def\bi{\begin{itemize}}
\def\ei{\end{itemize}}
\def\be{\begin{equation}}
 \def\ee{\end{equation}}
 \def\bea{\begin{eqnarray}}
 \def\eea{\end{eqnarray}}
 \def\bean{\begin{eqnarray*}}
 \def\eean{\end{eqnarray*}}
  \newcommand{\eg}{{\em e.g.}}

\newcommand{\eq}[1]{(\ref{#1})}

\newcommand{\mT}{M_\perp}
\newcommand{\thard}{t_{\mathrm{h}}}
\newcommand{\jpsi}{J/\psi}
\newcommand{\xf}{x_{\mathrm{F}}}
\newcommand{\pt}{p_{_\perp}}

\newcommand{\pp}{p--p}
\newcommand{\pA}{p--A}
\newcommand{\piA}{\ensuremath{\pi}--A}
\newcommand{\pN}{p--N}

\newcommand{\dd}{{\rm d}}
  
\newcommand{\lsim}{\lesssim} \newcommand{\gsim}{\gtrsim}

\newcommand{\qzero}{\hat{q}_0}
\newcommand{\gevsqfm}{GeV$^2$/fm}

\def\COMMENT#1{}

 \def\esim{\,\mathrel{\rlap{\lower0.2em\hbox{$-$}}\raise0.15em\hbox{\footnotesize $\hskip0.04em\sim$}}\,}
 \def\gsim{\mathrel{\rlap{\lower0.2em\hbox{$\sim$}}\raise0.2em\hbox{$>$}}}
 \def\ksim{\mathrel{\rlap{\lower0.2em\hbox{$\sim$}}\raise0.2em\hbox{$<$}}}

\def\xf{x_{_F}}

\def\pt{p_{_\perp}}

\begin{document}
\title{$\jpsi$ suppression in \pA\ collisions from parton energy loss in cold QCD matter }

\author{Fran\c{c}ois Arleo}
\affiliation{Laboratoire d'Annecy-le-Vieux de Physique Th\'eorique (LAPTh), UMR5108, Universit\'e de Savoie \& CNRS,  BP 110, 74941 Annecy-le-Vieux cedex, France}
\affiliation{CERN, PH-TH Dept. 1211 Geneva 23, Switzerland}
\author{St\'ephane Peign\'e}
\affiliation{SUBATECH, UMR 6457, Universit\'e de Nantes, Ecole des
Mines de Nantes, IN2P3/CNRS \\ 4 rue Alfred Kastler, 44307 Nantes cedex 3, France}

\date{\today}

\begin{abstract}
The effects of energy loss in cold nuclear matter on $\jpsi$ suppression in 
\pA\ 
collisions are studied. A simple model based on first principles and depending on a single free parameter is able to reproduce $\jpsi$ suppression data at large $\xf$ and at various center-of-mass energies. These results strongly support  energy loss as a dominant effect in quarkonium suppression. They also give some hint on its hadroproduction mechanism suggesting color neutralization to happen on long time-scales. Predictions for $\jpsi$ and $\Upsilon$ suppression in p--Pb collisions at the LHC are made.
\end{abstract}
\maketitle

\setcounter{footnote}{0}
\renewcommand{\thefootnote}{\arabic{footnote}}

The spectacular results on jet production in Pb--Pb collisions at the LHC (see e.g.~\cite{Collaboration:2010bu,Chatrchyan:2012ni}) find a natural explanation in terms of parton energy loss in quark-gluon plasma. Despite the wealth of data accumulated so far at RHIC and LHC, the in-depth understanding of energy loss processes remains far from complete (see~\cite{Armesto:2011ht} for a discussion). Recently, new scaling properties have been identified for the induced gluon radiation spectrum, $\dd I/\dd \omega$, of hard processes 
where a color charge undergoes small angle scattering through a static medium (nuclear matter or quark-gluon plasma) \cite{Arleo:2010rb}. 

The goal of this Letter is to explore the phenomenological consequences of these results on $\jpsi$ as well as $\Upsilon$ suppression,
\be
\label{RpA}
R_{\mathrm{pA}}\left(\xf\right) = {\frac{\dd\sigma_{\mathrm{pA}}^{\jpsi}}{\dd \xf} \left(\xf\right) \biggr/ \frac{A\ \dd\sigma_{\mathrm{pp}}^{\jpsi}}{\dd \xf} \left(\xf\right)} \, ,
\ee
in \pA\ and \piA\ collisions (we keep in the following the notations ``$\jpsi$'' and ``\pA'' for clarity). This observable allows for probing energy loss in nuclear matter, which is a well-controlled medium as opposed to an expanding quark-gluon plasma which dynamics is more complex. Moreover, understanding quarkonium suppression in \pA\ collisions is a prerequisite in order to interpret quantitatively the measurements performed in heavy-ion collisions at the LHC~\cite{Aad:2010aa,Chatrchyan:2012np,Abelev:2012rv}. It is striking that there is no consensus yet on the origin of the significant $\jpsi$ suppression reported at large rapidity in \pA\ collisions, from SPS to RHIC~\cite{Badier:1983dg,Leitch:1999ea,Adare:2010fn}, 
despite many theoretical attempts~(see~\cite{Frawley:2008kk} for a review). 

As is well-known, quarkonium hadroproduction in elementary \pp\ collisions is not fully understood. In order to study quarkonium nuclear suppression in the most model-independent way, the \pp\ quarkonium production cross section will be taken from experiment. We will only assume that the heavy-quark $Q \bar{Q}$ pair is produced in a compact {\em color octet} state, within the hard process time-scale $t_{\mathrm{h}}$, and remains color octet for a time much longer than $t_{\mathrm{h}}$. 
In quarkonium production models where color neutralization is a soft, non-perturbative process, this assumption holds at any $\xf$. In the Color Singlet Model, we expect this assumption to be founded at large enough $\xf$, where the gluon emission required for color neutralization is constrained to be semi-hard (or even softish) by energy conservation. 
With this assumption, at sufficiently large quarkonium energy $E$ in the target rest frame, quarkonium hadroproduction looks like small angle scattering of a color charge~\footnote{We stress that the spectrum \eq{our-spectrum} and associated average loss \eq{meanE} are absent in Drell-Yan production, where no color charge is produced at the production time $t_{\mathrm{h}}$.}. The associated soft gluon radiation spectrum is thus similar to the (non-abelian) Bethe-Heitler spectrum of an asymptotic charge, and depends on the amount of transverse momentum kick $q_\perp$ to the charge. The typical $q_\perp$ is expected to be larger in \pA\ than in \pp\ collisions due to transverse momentum {\em nuclear broadening} $\Delta q_\perp^2$. As a result, the {\em medium-induced} radiation spectrum is similar to the Bethe-Heitler spectrum with $q_\perp^2$ replaced by $\Delta q_\perp^2$ (see~\cite{Arleo:2010rb}),
\be
\label{our-spectrum}
\omega \frac{\dd I}{\dd \omega} = \frac{N_c \alpha_s}{\pi} \left\{ \ln{\left(1+\frac{\Delta q_\perp^2 E^2}{M_\perp^2 \omega^2}\right)} - \ln{\left(1+\frac{\Lambda^2 E^2}{M_\perp^2 \omega^2}\right)} \right\} \, .
\ee
In the following we shall take $\alpha_s = 0.5$, 
$\Lambda = \Lambda_{QCD} = 0.25 \,{\rm GeV}$, $\pt=1$~GeV in the transverse mass $M_\perp=\sqrt{M^2+\pt^2}$, and $M = 3$~GeV ($M = 9$~GeV) for the mass of a compact $c\bar{c}$ ($b\bar{b}$) pair.

This leads to an average medium-induced radiative loss scaling as the quarkonium energy, $\Delta E \propto E$. In the limit $\Lambda^2 \ll \Delta q_\perp^2 \ll M_\perp^2$ we have
\be
\Delta E \equiv \int_0^E \dd\omega \,\omega \frac{\dd I}{\dd \omega} \simeq  N_c \, \alpha_s \, \frac{\sqrt{\Delta q_\perp^2}}{M_\perp} \, E \, .
\label{meanE}
\ee
The scaling $\Delta E \propto E$ was first postulated in~\cite{Gavin:1991qk} (also revisited in~\cite{Kopeliovich:2005ym}) yet this assumption was not motivated and the parametric dependence on $L$ and $M$ arbitrary (and different from \eq{meanE}). In Ref.~\cite{Brodsky:1992nq}, a bound on medium-induced energy loss was derived, $\Delta E \lsim E^0$, but in a specific setup where the nuclear broadening of the final tagged particle was neglected. 

The starting point of the model consists in expressing the $\jpsi$ differential production cross section $\dd\sigma/\dd\xf$ in \pA\ collisions simply as that in \pp\ collisions, with a shift in $\xf$ accounting for the energy loss $\varepsilon$ incurred by the octet $c \bar{c}$ pair propagating through the nucleus,
\be
\label{eq:xspA}
\frac{1}{A}\frac{\dd\sigma_{\mathrm{pA}}^{\jpsi}}{\dd \xf}\left(\xf\right) = \int_0^{E_{\mathrm{p}} -E} \dd\varepsilon \,{\cal P}(\varepsilon) \, \frac{\dd\sigma_{\mathrm{pp}}^{\jpsi}}{\dd\xf} \left(\xf+\delta \xf(\varepsilon)\right)  ,
\ee
where $\xf$ is defined as the longitudinal momentum fraction between the $\jpsi$ and projectile proton in the c.m. frame of an elementary \pN\ collision (of energy $\sqrt{s}$).
In the limit $\sqrt{s} \gg m_{\mathrm{p}}$ (with $m_{\mathrm{p}}$ the proton mass),
it reads
\be
\label{xfofE}
\xf = \xf(E) = \frac{E}{E_{\mathrm{p}}} - \frac{E_{\mathrm{p}}}{E} \, \frac{M_\perp^2}{s} \, ,
\ee
where $E$ and $E_{\mathrm{p}} \simeq s/(2 m_{\mathrm{p}})$ are respectively the $c \bar{c}$ pair and projectile proton energies in the nucleus rest frame. We now describe the various ingredients in Eq.~\eq{eq:xspA}:\\
\noindent (i) The differential \pp\ cross section is determined from a fit of \pp\ data and can be conveniently parameterized as
\be
\label{pp-fit}
\frac{\dd\sigma_{\mathrm{pp}}^{\jpsi}}{\dd\xf} \left(\xf\right) \propto (1-x^\prime)^{n} / x^\prime \
;\quad x^\prime \equiv \sqrt{\xf^2+4 \mT^2/s}\ .
\ee
The exponent $n$ is extracted from \pp\ data taken at the same c.m. energy (whenever possible) as the \pA\ measurements discussed in this Letter (the normalization parameter being irrelevant 
for our purpose, see \eq{RpA});\\ 
\noindent (ii) The shift $\delta \xf(\varepsilon)$ is defined by
\be
\label{xfofEplusepsilon}
\xf(E)+\delta \xf(\varepsilon) = \xf(E+\varepsilon) = \frac{E+\varepsilon}{E_{\mathrm{p}}} - \frac{E_{\mathrm{p}}}{E+\varepsilon} \, \frac{M_\perp^2}{s} \, ,
\ee
where $E$ is obtained directly from (\ref{xfofE}). Note that at large $\xf\gg{M_\perp/\sqrt{s}}$, we have $E \simeq \xf E_{\mathrm{p}}$ and $\delta \xf(\varepsilon) \simeq \varepsilon / E_{\mathrm{p}}$;\\
\noindent (iii) The average over $\varepsilon$ in Eq.~(\ref{eq:xspA}) is performed using the energy loss probability distribution, or {\em quenching weight} ${\cal P}(\varepsilon)$. As a simple and physically sound choice for a normalized distribution, we take~\cite{ap} 
\be
\label{Pstar}
{\cal P}(\varepsilon) =  \frac{\dd I}{\dd\varepsilon} \, \exp \left\{ - \int_{\varepsilon}^{\infty} \dd\omega  \frac{\dd I}{\dd\omega} \right\} \, .
\ee

The amount of medium-induced gluon radiation, and hence the strength of $\jpsi$ suppression in \pA\ collisions, is controlled by $\Delta q_\perp^2$
in Eq.~\eq{our-spectrum}.
For a path length $L$ travelled across the target (proton or nucleus), it reads
\be
\Delta q_\perp^2(L) = \hat{q}_{\mathrm{A}} \, L - \hat{q}_{{\mathrm{p}}} \, L_{\mathrm{p}} \ ,
\label{broadening}
\ee
where the average path length is given by
$L = \frac{3}{2} \, r_0 \, A^{1/3}$ ($r_0=1.12$~fm), assuming the hard process to occur uniformly in the nuclear volume. In (\ref{broadening}), $\hat{q}_{\mathrm{A}}$ (resp. $\hat{q}_{\mathrm{p}}$) stands for the transport coefficient in the nucleus~A (resp. in the proton). It is related to the gluon distribution $G(x)$ in a target nucleon as~\cite{Baier:1996sk}
\be
\hat{q}(x) = \frac{4 \pi^2 \alpha_s(\hat{q} L) N_c}{N_c^2-1} \, \rho \, x G(x,\hat{q} L) \, ,
\label{qhat-gluondensity}
\ee
where $\rho$ is the target nuclear density. Apart from the scaling violations in the running of $\alpha_s$ and in the evolution of the gluon density, both neglected since $\hat{q} L\lesssim 1$~GeV$^2$, the $L$-dependence of $\hat{q}$ mainly enters via the typical $x$ value at which $xG(x)$ should be evaluated. When the hard production time $\thard \ll L$, it is estimated to be $x\simeq {(2 m_{\mathrm{p}} L)}^{-1}\equiv x_0$~\cite{Baier:1996sk}. When $\thard \gg L$,  the hard subprocess is coherent over the whole nucleus, and in this case we expect $x \sim x_2$~\cite{ap}, where $x_2 $ is the target parton momentum fraction, $x_2= (-\xf +x')/2$ when assuming a $2\to1$ subprocess kinematics. Using the power-law behavior $xG(x) \sim x^{-0.3}$ suggested by fits to HERA data~\cite{GolecBiernat:1998js}, $\hat{q}$ is thus given by~\footnote{Note that the $x$-dependence of $\hat{q}$ is not essential:  similar results  would be obtained using a constant $\hat{q}(x)=\hat{q}$.}
\be
\label{qhat-model}
\hat{q}(x) = \hat{q}_0 \left( \frac{10^{-2}}{x} \right)^{0.3}\ ;\ x = \min(x_0, x_2) \, .
\ee
In (\ref{qhat-model}), the transport coefficient $\hat{q}_0\equiv\hat{q}(x=10^{-2})$ is the {\em only free parameter} of the model. 

Besides energy loss effects, other mechanisms might affect $\jpsi$ suppression in nuclei. At small energy $E$, when the $\jpsi$ hadronization time $t_\psi=\gamma\tau_\psi \lesssim L$, $\jpsi$'s are produced inside the nuclear medium, and consequently might suffer inelastic interaction with the target nucleus, or nuclear absorption. In the figures below, we indicate by an arrow the typical value of $\xf$ below which this starts to happen, assuming $\tau_\psi \simeq0.3$~fm. Another effect is the expected saturation of the nuclear gluon density at small $x_2$, leading to an additional $\jpsi$ suppression in high-energy \pA\ collisions. %(with respect to \pp\ collisions). 
The 
associated
suppression is a scaling function of the saturation scale $Q_s$, which can be simply parameterized as~\cite{Fujii:2006ab}
\be
{\cal S}_{\mathrm A}^{\rm{sat}}(x_2, L) \simeq \frac{a}{\left(b+Q_s^2(x_2, L)\right)^\alpha}.
%R_{\mathrm{pA}}^{\rm{sat}}(x_2, A) \simeq \frac{a}{\left(b+Q_s^2(x_2, L)\right)^\alpha}.
\label{RpAsat}
\ee
In order to make reliable predictions at RHIC and LHC, where saturation effects might be important, the $\jpsi$ nuclear production ratio will be determined assuming energy loss effects, $R_{\mathrm{pA}}^{\rm{E.loss}}$ from Eq.~(\ref{eq:xspA}), with and without saturation effects,
\bea
{\rm (i)}\qquad R_{\mathrm{pA}} &=& R_{\mathrm{pA}}^{\rm{E.loss}}\ ,\nonumber\\
{\rm (ii)}\qquad R_{\mathrm{pA}} &=& R_{\mathrm{pA}}^{\rm{E.loss}} \times {\cal S}_{\mathrm A}^{\rm{sat}} / {\cal S}_{\mathrm p}^{\rm{sat}}\ . \nonumber
\eea
The saturation scale appearing in (\ref{RpAsat}) is determined consistently in our model through the relationship~\cite{Mueller:1999wm}
\be
\label{Qs-gluondensity}
Q_s^2(x, L) = \hat{q}(x) \, L,
\ee
where $\hat{q}(x)$ is given by (\ref{qhat-model}). The inclusion of saturation effects thus does not require any additional parameter.

%%%%%%%%%%%%%%%%%%%%%%%%%%%%%%%%%%%%%%%%%%
\begin{figure}[htbp]
\begin{center}
    \includegraphics[width=8.8cm,height=5.4cm]{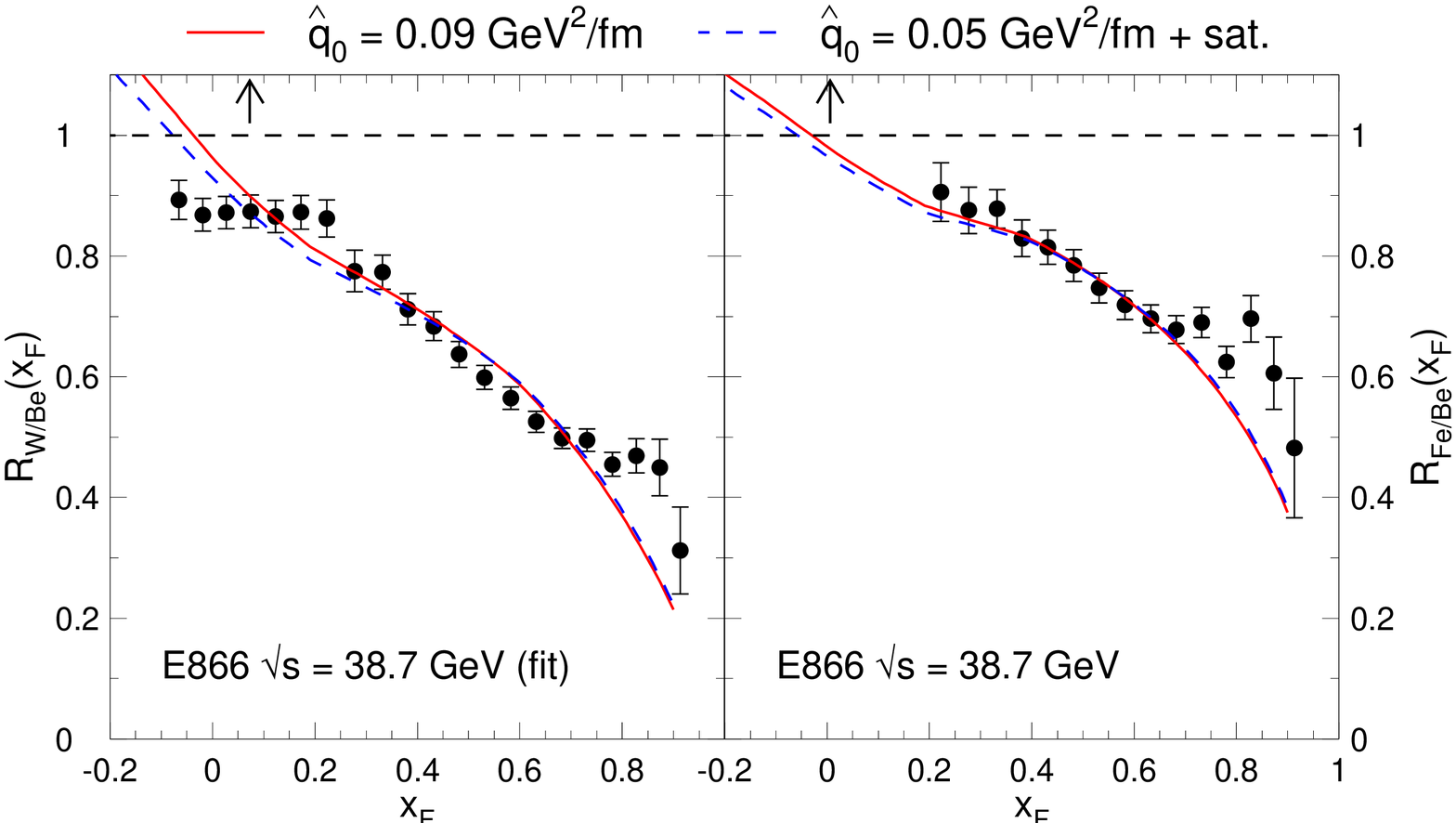}
  \end{center}
\caption{E866 $\jpsi$ suppression data~\cite{Leitch:1999ea} in p--W (left) and p--Fe (right) collisions compared to the energy loss model.}
  \label{fig:e866}
\end{figure}
%%%%%%%%%%%%%%%%%%%%%%%%%%%%%%%%%%%%%%%%%%
The only parameter of the model, the transport coefficient $\qzero$, is determined by fitting the $\jpsi$ suppression measured by E866~\cite{Leitch:1999ea} in p--W over p--Be collisions at $\sqrt{s}=38.7$~GeV in the [0.2--0.8] $\xf$-range~\footnote{Depending on the $\xf$ range used for the fit, the resulting uncertainty on $\qzero$ is of the order of 20\%.}. This choice is motivated by the fact that these data are the most precise measurements performed so far and cover a wide range in $\xf$. The fit gives $\qzero=0.09$~\gevsqfm\ assuming energy loss effects only, and $\qzero=0.05$~\gevsqfm\ when saturation effects are also included. The result of the fit 
in these two cases is shown in Fig.~\ref{fig:e866} (left panel); the agreement is excellent in the fit range, while a slight disagreement is observed below $\xf\lesssim0.1$, where nuclear absorption is expected to play a role. The successful description of $\jpsi$ suppression in iron, $R_{\rm{Fe/Be}}$ (right panel), at the same energy fully supports the atomic mass dependence of the model. Note also that the values for the transport coefficient, $\qzero=0.05$--$0.09$~\gevsqfm, would correspond to $Q_s^2(x=10^{-2})=0.08$--$0.15 \ \rm{GeV}^2$ using (\ref{Qs-gluondensity}), which is consistent with (yet slightly smaller than) estimates based from fits to $F_2$ DIS data~\cite{Albacete:2010sy}.

%%%%%%%%%%%%%%%%%%%%%%%%%%%%%%%%%%%%%%%%%%
\begin{figure}[htbp]
  \begin{center}
    \includegraphics[width=8.8cm,height=8.cm]{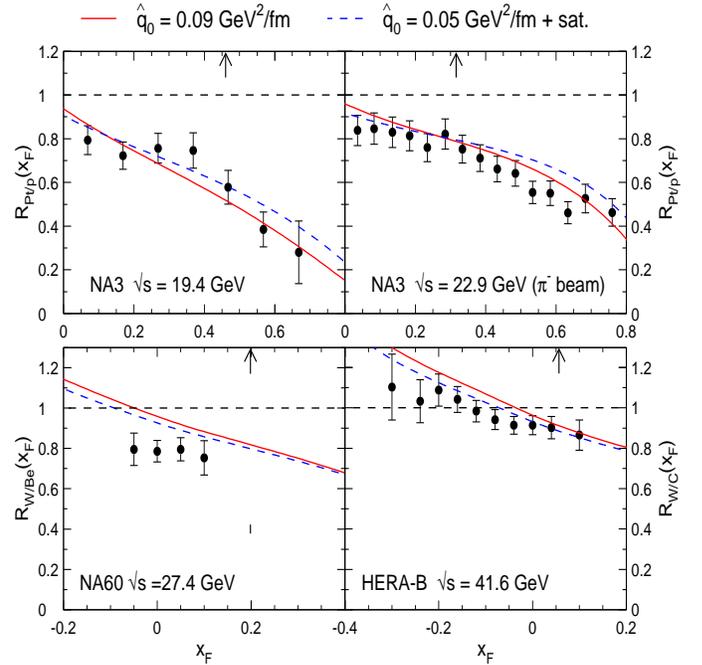}
  \end{center}
\caption{NA3~\cite{Badier:1983dg}, NA60~\cite{Arnaldi:2010ky} and HERA-B~\cite{Abt:2008ya} $\jpsi$ suppression data in p--A ($\pi^-$--A) collisions compared to the energy loss model.}
  \label{fig:lowenergy}
\end{figure}
%%%%%%%%%%%%%%%%%%%%%%%%%%%%%%%%%%%%%%%%%%
Data taken at lower $\sqrt{s}$ or smaller $\xf$ are also compared to the model. As can be seen in Fig.~\ref{fig:lowenergy} the agreement is very good, both in shape and magnitude, over a very wide range in $\xf$. We also remark that as expected, the model tends to slightly underestimate the suppression in the $\xf$ domain where $\jpsi$'s might hadronize inside the nucleus and suffer nuclear absorption; see in particular the comparison with NA60~\cite{Arnaldi:2010ky} and HERA-B~\cite{Abt:2008ya} data. It is also remarkable that the model is able to reproduce the different magnitude of suppression in \pA\ and $\pi^-$--A collisions reported by NA3~\cite{Badier:1983dg} (Fig.~\ref{fig:lowenergy}, upper panels). This difference cannot be understood within nuclear absorption models (since factorization is usually assumed between the hard production process and final state interaction) nor can it be explained by nuclear PDF (nPDF) effects, unless the nPDF 
to proton PDF ratios for valence quarks and for gluons, probed respectively in $\pi^-$--A and \pA\ collisions, prove completely different. In our picture, the smaller $\jpsi$ suppression in $\pi^-$--A collisions naturally arises from the flatter differential cross section, $n_{\pi{{\mathrm{p}}}}=1.5$ vs. $n_{{\mathrm{p}}{\mathrm{p}}}=4.3$ (see~(\ref{pp-fit})) at NA3 energies, a feature which can be explained from the slope of the PDF in a pion and in a proton, respectively. For completeness, the predictions including saturation are also shown as dashed lines in~Fig.~\ref{fig:lowenergy}; as expected saturation effects are tiny at these energies. The slight differences are essentially due to the smaller transport coefficient used ($\qzero=0.05$~\gevsqfm) when saturation is included.

%%%%%%%%%%%%%%%%%%%%%%%%%%%%%%%%%%%%%%%%%%
\begin{figure}[htbp]
  \begin{center}
    \includegraphics[width=5.8cm]{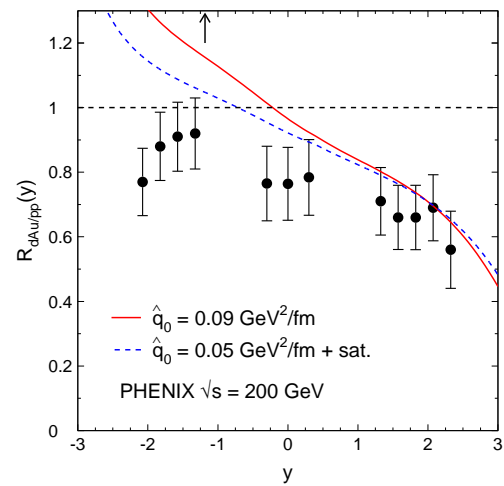}
  \end{center}
\caption{PHENIX $\jpsi$ suppression data~\cite{Adare:2010fn} in d--Au collisions compared to the energy loss model.}
  \label{fig:rhic}
\end{figure}
%%%%%%%%%%%%%%%%%%%%%%%%%%%%%%%%%%%%%%%%%%
The predictions in d--Au collisions at RHIC, $\sqrt{s}=200$~GeV, are shown in Fig.~\ref{fig:rhic} in comparison with PHENIX data~\cite{Adare:2010fn}. Energy loss effects are able to reproduce $\jpsi$ suppression at positive rapidities. However, some disagreement is observed in the negative $y$ bins 
for which nuclear absorption is expected to play a role; it has also been conjectured that a depletion of the gluon nPDF at large $x_2$ might explain the trend of the data at backward rapidities~\cite{Ferreiro:2011xy}.  We also note that the disagreement is reduced when saturation effects are included. Finally, arguing about the possible slight disagreement observed around mid-rapidity might be premature, given the present experimental uncertainties.

%%%%%%%%%%%%%%%%%%%%%%%%%%%%%%%%%%%%%%%%%%
\begin{figure}[htbp]
  \begin{center}
    \includegraphics[width=5.8cm]{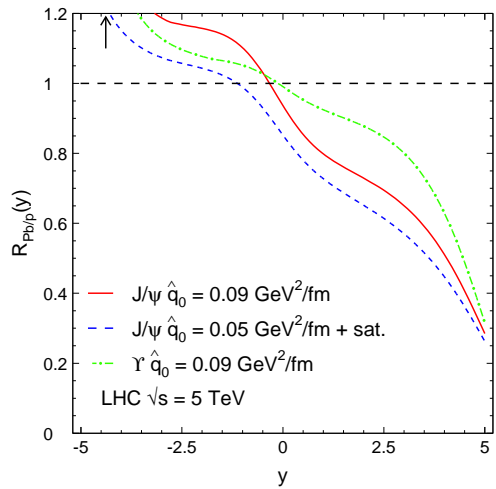}
  \end{center}
\caption{$\jpsi$ and $\Upsilon$ suppression expected in p--Pb collisions at the LHC.}
  \label{fig:lhc}
\end{figure}
%%%%%%%%%%%%%%%%%%%%%%%%%%%%%%%%%%%%%%%%%%
Last, the rapidity dependence of $\jpsi$ suppression in p--Pb collisions at the LHC (taking $\sqrt{s}=5$~TeV) is shown in Fig.~\ref{fig:lhc}.
Even though the inclusion of saturation effects is expected to yield a stronger $\jpsi$ suppression, it is somehow compensated by the use of a smaller transport coefficient; as a consequence predictions with (dashed line) and without (solid) saturation are actually rather similar, except in the negative $y$-bins. As can be seen from the arrow in Fig.~\ref{fig:lhc}, $\jpsi$ hadronization should take place outside the nuclear medium above $y\gtrsim-5$; nuclear absorption should thus play little or no role at the LHC. At forward rapidities, $\jpsi$ suppression becomes rather large, \eg, $R_{\rm{pPb}}\simeq0.7$--$0.8$ at $y=1$ down to $R_{\rm{pPb}}\lsim0.5$ at $y\gtrsim4$. 
Fig.~\ref{fig:lhc} also shows the predicted $\Upsilon$ suppression as a dash-dotted line~\footnote{Lacking $\Upsilon$ p--p data, we assume $n^\Upsilon=n^{\jpsi}$ in Eq.~(\ref{pp-fit}).}. Because of the mass dependence of energy loss, $\Delta E \propto {\mT}^{-1}$, it is expected to be smaller than that of $\jpsi$ yet not negligible, \eg, $R^{\Upsilon}_{\rm{pPb}}\simeq0.8$ at $y=3$. These predictions can be compared to the future measurements by the ALICE and LHCb experiments during the p--Pb run scheduled in 2012.

In summary, an energy loss model (supplemented by saturation effects) based on first principles has been presented. Once the transport coefficient is determined from a subset of E866 data, it is able to reproduce nicely all existing $\jpsi$ measurements in p--A collisions. In particular the dependence of $\jpsi$ suppression on $\xf$, for various atomic mass $A$ and center-of-mass energies $\sqrt{s}$, is well accounted by the model. These results strongly support energy loss in cold nuclear matter to be the dominant effect of $\jpsi$ suppression, at least at large rapidities. It also qualitatively explains why no suppression in inelastic $\jpsi$ electroproduction is observed~\cite{Amaudruz:1991sr}, since Eq.~(\ref{meanE}) does not apply to this case~\cite{Arleo:2010rb}. Finally, the agreement between the data and our model predictions supports the assumption of a long-lived color octet $Q \bar{Q}$ pair. As a perspective, we plan to investigate energy loss effects on $\jpsi$ suppression in heavy-ion collisions as well as to explore their consequences on other hard processes.

\acknowledgments 
We would like to thank S.~Brodsky, Y.~Dokshitzer, D.~Kharzeev, B.~Kopeliovich and M.~Strikman for useful exchanges. FA thanks CERN PH-TH division for hospitality. This work is funded by ``Agence Nationale de la Recherche'' under grant ANR-PARTONPROP.

\providecommand{\href}[2]{#2}\begingroup\raggedright\endgroup

\end{document}